# Secure and Resilient Low-Rate Connectivity for Smart Energy Applications through Power Talk in DC Microgrids

Čedomir Stefanović, Marko Angjelichinoski, Pietro Danzi, and Petar Popovski[1]

**ABSTRACT**

The future smart grid is envisioned as a network of interconnected microgrids (MGs) – small-scale local power networks comprising generators, storage capacities and loads. MGs bring unprecedented modularity, efficiency, sustainability, and resilience to the power grid as a whole. Due to the high share of renewable generation, MGs require innovative concepts for control and optimization, giving rise to a novel class of smart energy applications, in which communications represent an integral part. In this paper, we review *power talk*, a communication technique specifically developed for direct current MGs, which exploits the communication potential residing within the MG power equipment. Depending on the smart energy application, power talk can be used either as a primary communication enabler, or an auxiliary communication system that provides resilient and secure operation. The key advantage of power talk is that it derives its availability, reliability, and security from the very MG elements, outmatching standard, off-the shelf communication solutions.

## 1. INTRODUCTION

The architecture of the power grid has been experiencing a paradigm shift from the classic organization in bulk-generation, transmission and distribution subsystems into a flexible structure with a high penetration of microgrids (MGs), see Fig 1. MG is a localized collection of distributed energy resources (DERs), storages and loads, operating connected to the main grid or in a standalone, islanded mode [1]. The goal is to achieve self-sustainable, efficient and resilient operation, thereby improving the operation of the entire power network.

Recent advances show that MGs are becoming economically viable [2]. In particular, direct current (DC) MGs are gaining popularity due the fact that most of renewable DERs, storages and modern loads are DC in nature, implying simpler implementation, reduced costs, higher efficiency and increased resilience to the main grid disturbances with respect to AC MGs. However, there are several challenges yet to be solved to foster large scale implementation of DC MGs. From the research and development angle, the major challenge is to incorporate control and communication features pertinent to DC MGs [2,3]. Specifically, renewable DERs show high unpredictability and variability in comparison to traditional bulk generation, requiring novel control and optimization approaches, both at the intra- and inter-MG level.

DC MG control architecture is organized in primary, secondary, and tertiary layers [3]. The primary control is the fastest, operating in the frequency range 0.1 – 1 MHz, regulating the MG voltage and power flow such that high frequency load/generation variations are compensated. It is implemented in a decentralized manner, and it uses the measurements locally available to DERs and the control references provided by the tertiary control. It. The secondary control operates using frequency 1 – 10 kHz, its task is to eliminate the steady state voltage drift and power sharing mismatch, introduced by the primary control due to its decentralized nature. Finally, the tertiary control comprises smart energy applications that minimize power dissipation losses and generation costs, as well as maximize the economic viability of the system by providing the optimal references for the primary control. It is the slowest control level, running periodically every 5-30 minutes. Typical smart energy applications include Optimal Power Flow (OPF), Optimal Economic Dispatch (OED), Demand-Response (DR), Unit Commitment (UC), etc.

---

[1] The authors are with the Department of Electronic Systems, Aalborg University, Denmark.

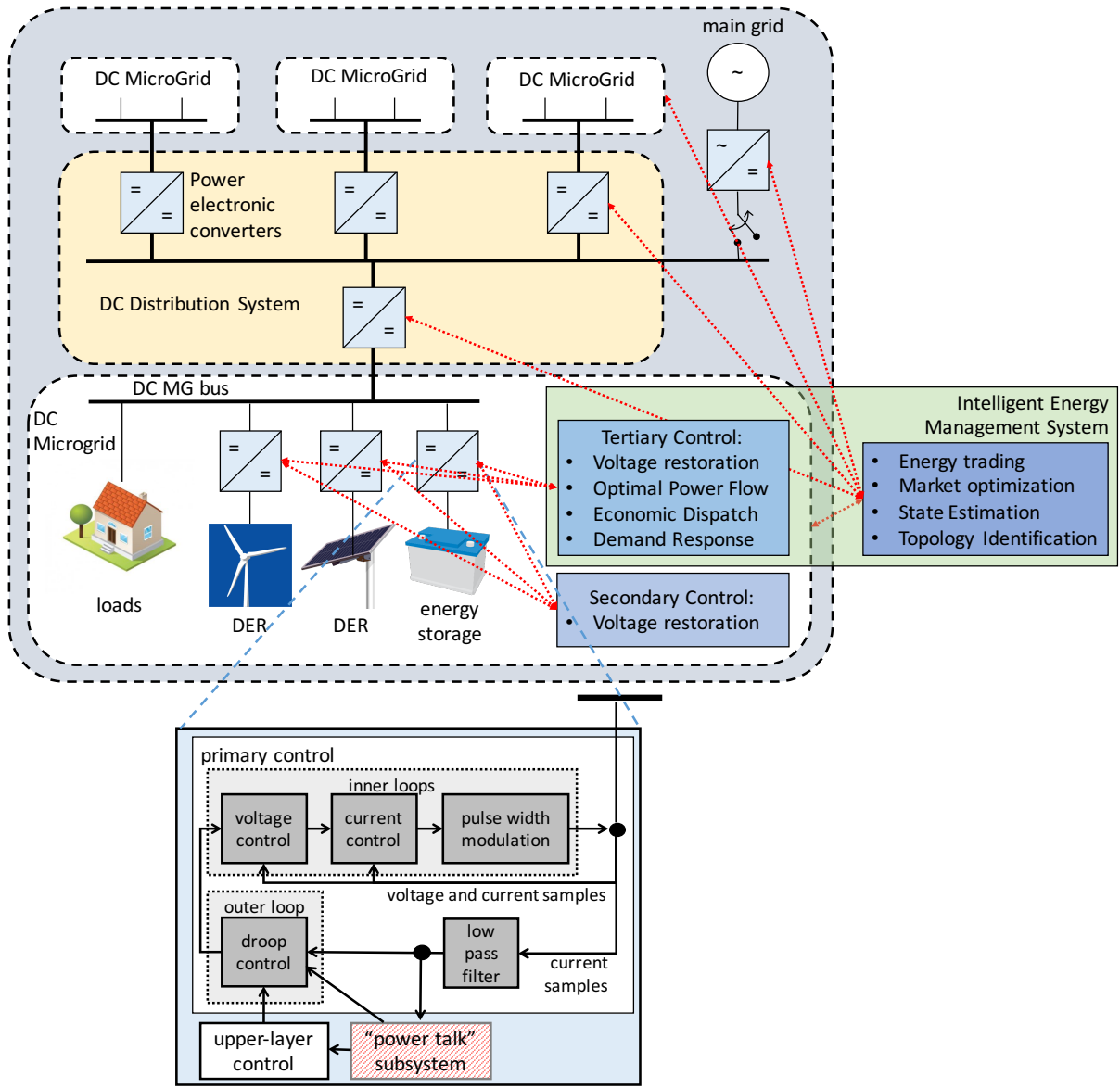

*Figure 1: Cluster of DC microgrids: DC MGs are characterized by extensive use of power electronic converters that regulate the power generation, executing intra- and inter-MG control algorithms.*

Unlike the primary control layer, the secondary and tertiary control layers require information that is not available via local measurements. Depending on the control application, this information may comprise voltage/current measurement at remote DERs, instant DER generation capacities, loads' demands and admittance matrix, control directives, information on generation costs, ramp-up constraints etc. [3]. In other words, secondary and tertiary MG control applications require communication support. The actual amount of data that should be communicated is small, as typical for machine-to-machine communications, and the periodicity of the communication exchanges should follow the periodicity (i.e., frequency) of the control application.

The information exchange is also required to enable higher level inter-MGs operation, executed within and among MG clusters, see Fig. 1. In fact, modern MGs have the capability of integrating with the existing power system and dealing with bidirectional exchange of power, thus increasing the overall grid availability in case of fault events and reducing the stress on overloaded portions of the system [4]. Moreover, the interaction of PECs running higher level control applications enables state estimation, topology identification, energy trading, and market optimization. These higher layer MG control applications, also including tertiary MG control, form an Intelligent Energy

Management System (IEMS), see Fig. 1, which supervises and optimizes the overall MG operation in a broader environment where the MG is placed [5], like commercial buildings and residential blocks.

To meet the communication needs of intra- and inter-MG control and optimization, a standard approach is to employ an external communication solution, such as wireless or powerline communications (PLC) [3,4,6]. However, relying only on external communication systems may compromise the goal of self-sustainable and resilient MG operation, due to their limited availability, reliability and security, cf. [7]. In this paper, we propose a communication framework for DC MGs that exploits *only* the communication potential residing within the MG power equipment. Depending on the smart energy application, power talk can be used either as a primary communication enabler, or an auxiliary communication system that fosters resilient and secure operation. Besides the fact that power talk derives its availability, reliability, and security from the very MG components, it also does not require installation of any additional hardware and provides complete coverage over the MG system.

## 2. BASICS OF POWER TALK

MGs are characterized by an extensive use of power electronic converters (PECs), which are digital signal processors interfacing DERs and storages to the buses [2,3], see Fig. 1. In a DC MG, a PEC (i.e., the DER it controls) can operate either as a voltage source converter (VSC) or a current source converter (CSC). When operating in VSC mode, PEC participates in MG control and optimization through regulation of the output power of its DER. In CSC mode, PEC does not participate in MG control and optimization, and the DER it controls generates the maximum output power. Note that a PEC can change its operating mode from VSC to CSC and vice versa as needed.

The block diagram of a PEC operating as a VSC can be seen in Fig. 1. A VSC constantly measures bus voltage and current with switching frequency that ranges from several tens of kHz to couple of MHz, and, based on the measurements, executes the primary control algorithm that is standardly implemented in the form of droop control [3]. VSCs also perform the upper layer control and optimization functions, which require communication support.

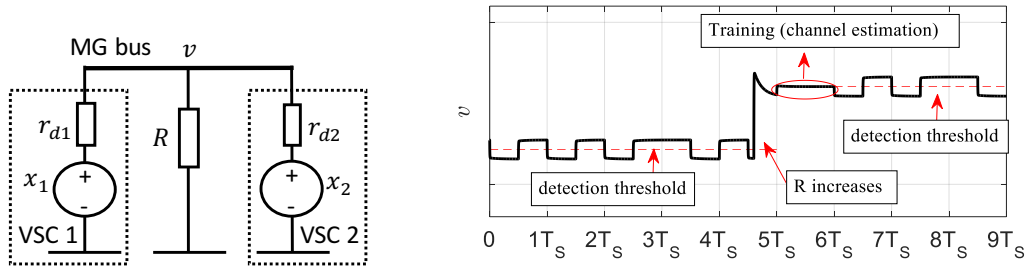

*Figure 2: (a) Example of a single-bus MG. (b) Communication from VSC 1 to VSC 2 via bus voltage deviations and the effects of the load change.*

Being digital signal processors, PECs can also engage in communication-related tasks using the buses interconnecting them as the communication medium. We illustrate this through a simple example of a single-bus MG, depicted in Fig. 2, with two PECs that operate as droop controlled VSCs and a single resistive load $R$. The steady-state bus voltage $v$, observed by both VSCs, is given by:

$$v = \frac{R(r_{d1} x_2 + r_{d2} x_1)}{R(r_{d1}+r_{d2}) + r_{d1} r_{d2}} \qquad (1)$$

where $x_1/x_2$ is the nominal reference voltage and $r_{d1}/r_{d2}$ is the virtual resistance of VSC1/VSC2, which are controllable droop parameters. Obviously, if VSC1 deviates its $x_1$ and/or $r_{d1}$, this will cause deviations of $v$. From the communication engineering point of view, $x_1$ and $r_{d1}$ can be seen as the *inputs*, and $v$ as the *output* of the communication channel between VSC1 and VSC2. This simple, but fundamental insight can be exploited to design of a communication system among PECs in the MG, which is embedded in the primary control and which uses steady-state deviations of the bus voltages to transfer information. This is the key concept of *power talk*, which can be implemented via a software modification of PEC architecture.

The idea of using MG bus for communication among PECs was proposed in several works that target predefined MG setups and control applications, in which PECs perform control actions based on the observation of the bus voltage, cf. [8] and the references therein. Establishment of a low-rate communication interface over DC bus by selecting predefined PEC switching frequencies was proposed in [9], and using pulse-width modulation in [10]. Both [9] and [10] address only physical layer aspects, neglecting the functionalities needed for setting-up fully operational communication links. In contrast, power talk is designed with an aim to establish a general digital interface among PECs in MG that can be used for any control and optimization application. Finally, we remark that power talk uses power lines to convey information like in PLC. However, all PLC standards require installation of dedicated communication hardware, whereas power talk is envisioned as an upgrade of the control functionality of PECs with communication capabilities, without using any additional communication hardware.

## MAIN CHARACTERISTICS OF POWER TALK

To create a functional communication solution through primary control, there are several important aspects to be taken account. We elaborate them in a general multibus MG setup, see Fig. 3.

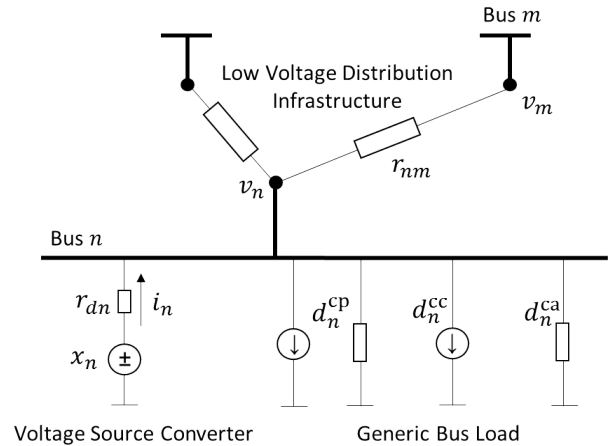

*Figure 3: General electrical model of a multibus MG. Per bus, there is a single VSC and an aggregate load representation with a constant power, a constant current and a resistive component, with power demands $d^{cp}$, $d^{cc}$ and $d^{cp}$ at rated bus-voltage $x_R$, respectively; the constant power component of the load also accounts for the potential CSCs in the bus. $r_{nm}$ denotes the resistance of the line connecting buses $n$ and $m$.*

*Signaling rate*: A MG bus requires typically $1 - 10$ ms to reach a steady state, implying that signaling rates in power talk are of the order of $100 - 1$ kBd. These rates are adequate for all IEMS applications, but not for secondary control.

*Synchronization*: Virtually all MG control applications have a periodic nature, such that power talk should be invoked in regular intervals. Further, being a baseband digital technique, power talk also requires packet- and symbol-level synchronization. If PECs are equipped with an external synchronization interface, like GPS, the synchronization may be easily achieved. Otherwise, PECs can rely on their internal clocks for coarse synchronization, and then apply standard techniques to achieve and maintain precise packet- and symbol-level synchronization, e.g., use of synchronization preambles, adequate signaling formats [11], etc.

*Multiple access*: Power talk establishes multiple access communication channels. This is readily observed from Eq. (1), which shows that the steady-state voltage of the bus depends on the control parameters of *all* VSCs in the MG. Considering the expected periodicity of power talk sessions and static/slowly changing MG control configuration (i.e., assignment of PEC operating modes), a straightforward approach is to employ time-division multiplex [12], essentially creating half-duplex channels. Another appealing option, motivated by the fact the set of transmitters is a priori known, is to use coding strategies for multiple access. In this approach, all VSCs simultaneously exchange information among themselves [12], achieving all-to-all full duplex communications.

*Channel state information*: Application of Kirchoff's laws reveals that the steady-state bus voltage $v_n$ that VSC n observes (see Fig. 3) depends on values of *all* reference voltages $x$, virtual resistances $r_d$, line resistances $r$ and load components $\mathrm{d^{cp}}$, $\mathrm{d^{cc}}$ and $\mathrm{d^{cp}}$ in the system [12]. In other words, the state of the communication channel is determined by the values of all components of the electrical model of the MG. The knowledge of these values is a priori unavailable, necessitating a training phase in which receiving VSCs learn the states of the channels they observe before engaging in communications. The training can be done through coordinated actions of all VSCs in the system [12]. This is reminiscent of standard approaches in wireless communication systems where the channels are subject to random behavior and some form of training is required for channel estimation.

*Load changes*: Loads change in MGs change sporadically, but unpredictably. If a load change occurs during a power talk session, the channel state information becomes invalidated, which may require restart of training phase and of information transfer, as illustrated in Fig 2(b). Load changes may be detected using standard error detection methods on physical layer, e.g., using CRC codes.

*Noise*: The observations of bus voltages contain measurement noise, which can be modelled as additive, Gaussian and white, where the typical values for the standard deviation of the voltage measurement noise (in volts/sample) in low voltage distribution systems is in the range of 0.01 – 0.1 % of the voltage rating per unit [12]. The SNR of power talk is determined by the amplitude of the allowed bus-voltage deviations used for power talk that the MG can tolerate, and the noise power after averaging of the samples during a signaling interval. Nevertheless, it can be shown that power talk operates in a very high SNR regime [12], such that the impact of noise is rather small. Finally, if required, the impact of noise can be further reduced using channel coding methods.

*Electrical constraints*: Virtual resistances and reference voltages by default feature constraints on their minimal and maximal value. Moreover, information-carrying deviations of these control parameters incur power deviations on the buses, and should be as small as possible with respect to the optimal power levels prescribed by the smart energy applications. In terms of communication system design, these constraints define the *signaling space* of power talk, in which one can construct optimized symbol constellations [13].

*Security*: Power talk, like PLC in general, offers security advantages with respect to the use of wireless networking for the MG information exchange. It is robust against cyber-attacks, e.g. information confidentiality, as an eavesdropper must have access to the physical infrastructure of the MG, as well as robust against both unintentional and intentional electromagnetic interference. With respect to PLC, power talk has the advantage that the communication is directly actuated by the PEC control software without delegating it to an external modem. In contrast, PLC require establishment of a trustful relationship of the control layer with the external communication network.

## 3. THE ARCHITECTURE OF SMART ENERGY APPLICATIONS WITH POWER TALK

The proposed functional architecture of PECs executing smart energy applications with power talk is depicted in Fig. 3. The power talk block provides information exchanges for the IEMS, i.e., for all tertiary and inter-MG control and optimization applications, as these have slow dynamics and require modest data rates. On the other hand, the frequency with which the secondary control operates is beyond reach of power talk, mandating use of an external, high-rate communication interface. Nevertheless, the power talk interface can be used as an auxiliary channel for the exchange of the information related to the status of the external network, such as connectivity status and alarm messages that can be used for the external network reconfiguration [14], or for establishment of a security context that can be used by the external network [15].

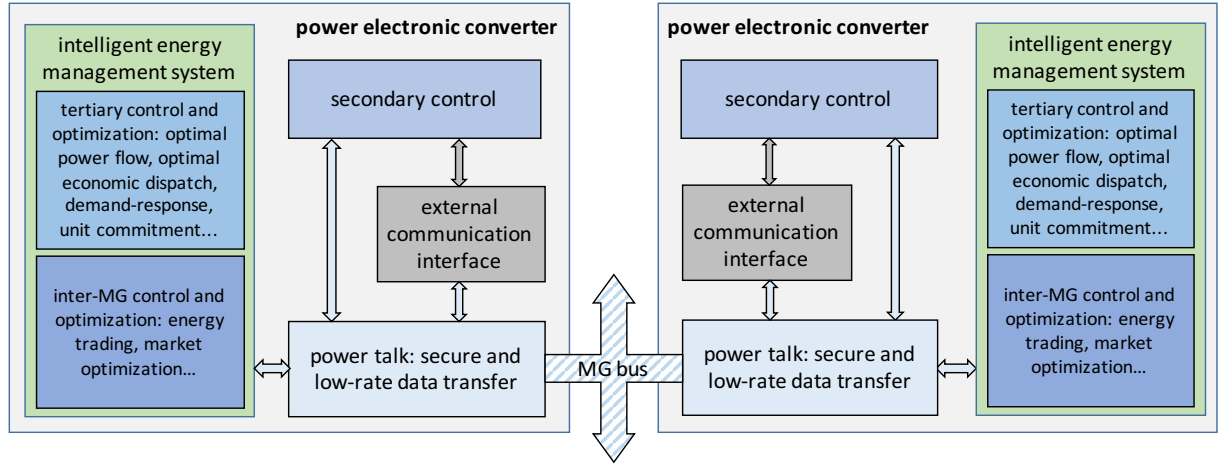

*Figure 4: The proposed functional architecture of PEC executing smart energy applications with power talk.*

In the following, we illustrate the potential of power talk via two example case studies: in Section 4, power talk is used as a communication solution for the tertiary control application of the optimal economic dispatch, while in Section 5, power talk is used in the context of the distributed secondary control as an auxiliary channel for the reconfiguration of the external wireless network under jamming attack.

## 4. CASE STUDY 1: OPTIMAL ECONOMIC DISPATCH

In MGs that are predominantly based on renewable technologies, the IEMS collects information about the generation capacities and runs OED periodically, e.g., every 5–30 minutes. The goal of OED is to dispatch the VSCs based on the instant generation capacities, such that the total generation cost is minimized and the load is balanced.

We focus on distributed OED (DOED) implementation with linear cost functions, typically used for renewable generation The MG hosts $U$ dispatchable VSCs, with generation capacities denoted by $p_{u,\max}$. Each VSC is assigned incremental cost $c_u$ per unit of generated power; without loss of generality the costs follow the ordering $c_1 < c_2 < \cdots < c_U$. The load demand is denoted with $d$. We assume a typical situation where $d$ is known a priori, via, e.g., an accurate forecasting performed one day in advance. In the distributed implementation, VSC $u$, besides $d$, needs to know $p_{k,\max}$ for each $k$ that satisfies $c_k < c_u$.

We design simple power talk protocol to support DOED. The protocol consists of periodic power talk phases, each phase preceding the next DOED period, during which DERs exchange the information required by DOED. A power talk phase consists of $N$ time slots of duration $T_S$ seconds. The multiple access in the power talk phase is via time division multiplex: the slots are divided into $U$ consecutive sub-phases, each sub-phase is assigned to one of the VSCs and consists of $Q$ slots, such that $N = QU$, as depicted n Fig. 5(a). VSC $u$ quantizes its generation capacity $p_{u,\max}$ into binary string of $Q$ bits, which is then transmitted in the dedicated sub-phase via uncoded binary modulation of the reference voltage $x_u$. Specifically, a logical 0/1 is transmitted by deviating the $x_u$ from its nominal value by a predefined level $-\gamma/\gamma$, which causes deviations of the bus voltage. The receiving VSCs simply compare the bus-voltage level observed in each slot to the bus-voltage level prior to the power talk phase, thereby detecting the transmitted bits.

At the end of the power talk phase, each VSC acquires the knowledge of the instant generation capacities. However, this knowledge is imperfect due to quantization and noise induced detection errors. Thus, the resulting dispatch policy in the next DOED period might be suboptimal, leading to an increase of the generation cost in comparison to the optimal policy. Moreover, the cost of the power deviations due to information-carrying bus-voltage deviations in the power talk phase should be also accounted for. These two factors represent the cost incurred by power talk.

We instantiate the proposed protocol in a single-bus MG with $U = 6$ VSCs, with the rated voltage of the bus is $x_R = 48$ V, the sampling (switching) frequency is 50 kHz, the standard deviation of the converters sampling noise is 0.05 volts/sample, and the duration of power talk slots is $T_S = 5$ ms. The samples obtained in each slot are averaged, based and value of the bits is decided. We measure the efficiency of the proposed approach via the average relative increase of the generation cost when power talk is used with respect to the generation cost when an ideal, "costless" communication solution is used, denoted by $\delta$. Fig. 5(b) depicts $\delta$ as function of the number of quantization bits $Q$, parameterized with reference voltage deviation amplitudes $\gamma$ in the range $0.02 - 1$ volts (corresponding to average power deviation in the range $5 - 200$ watts). We see that the largest values of $\delta$ occur for very small values of $Q$, as in this case the received information about the generation capacities is very imprecise. On the other hand, for $Q > 5$ the generation cost increase is becoming dominated by the power spent on the power talk phase. In this example, $\delta$ is minimized for $Q = 4$, proving that the length of the messages in smart energy applications is indeed very short. Finally, we note that the minimal relative cost increase is below 1%, making power talk a viable candidate in comparison to solutions that employ external communication systems, which involve costs of installation, maintenance and operation.

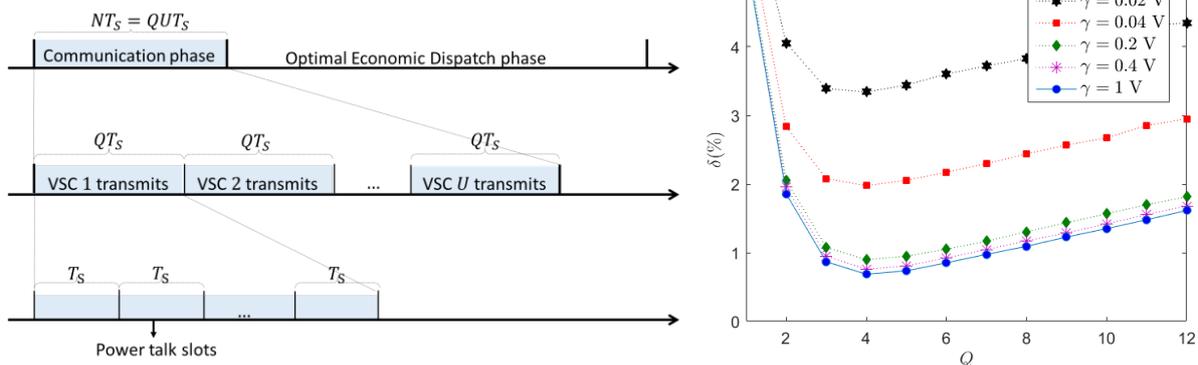

*Figure 5: (a) Temporal organization of the proposed protocol. (b) The relative cost increase when power talk is used in the communication phase of OED, compared to the ideal, costless communication solution.*

## 5. CASE STUDY 2: ROBUST AND SECURE DISTRIBUTED SECONDARY CONTROL

An envisioned application for low voltage MGs is the enforcement of the power reliability of critical buildings, such as commercial buildings, in which unexpected voltage fluctuations may cause damages to the electronic equipment. In this scenario, the MG is composed by a high number of small DERs that are networked by short range wireless communication interfaces, such as IEEE 802.11, to reduce the installation and operational costs, as depicted in Fig. 6(a). The voltage restoration is supported by distributed algorithms, that, contrary to the centralized control approach, permit an easy network reconfiguration, enhancing the grid scalability and relieving it from single point of failure [3]. The secondary control is executed by a subset of active DERs (i.e., VSC units), while the others work as CSCs in order to maximize the overall generation.

The communication graph of the networked VSCs should be strongly connected to enable proper execution of the secondary control. However, adverse channel conditions, such as continuous jamming, may cause the graph disconnection and the formation of insulated subsets of VSCs. In this case, a consensus-based secondary control is prevented to converge to a global solution, reflecting to the physical effect of unbalanced power sharing among DERs [7].

A possible approach to deal with such scenario is to select of a new subset of voltage regulators, by switching some of the CSCs to VSC operation mode such that the communication graph becomes connected again, while the insulated VSCs are switched to CSC mode. The proposed network reorganization can be done via periodically invoked power talk sessions, similarly to the approach outlined in Section 4. Specifically, the proposed protocol adopts the following steps: (i) all DERs broadcast wireless packets, (ii) based on the received wireless broadcasts, each DER broadcasts list of reachable neighbors and its current power generation capacity over the power talk channel in a TDMA fashion, where it is assumed that CSCs temporarily switch to VSC mode in order to participate in power talk communication, (iii) finally, each DER locally decides on its operation mode – VSC or CSC, such that the communications graph of the wireless network is connected, the voltage restored and the power sharing balanced. The last step is possible due to the fact all DERs share the same knowledge used for that purpose. In an unfavorable case in which jamming is such that it is not possible to wirelessly network VSCs to facilitate an adequate voltage restoration, the use of power talk enables dissemination of the information that can be used to detect this event.

We instantiate the proposed approach via the example depicted in Fig. 6(a), simulating in Simulink/PLECS a MG with the rated of voltage 48 V, composed by 9 DERs in which DERs 2,5,6 and 9 are initially participating in the secondary control. Their connectivity is undermined by a jamming device placed in proximity of DER 5, that is able of continuous transmission and prevents its communication. When a load variation occurs, as the activation of a resistive load at $t = 7$ s in Fig. 6(b), the absence of global connectivity reflects in a current imbalance. Nevertheless, the condition is detected in the consecutive power talk phase and a reorganization is triggered. The result is the formation of a new secondary control set, composed by DERs 1, 2, 7 and 10. In the simulation we adopted IEEE 802.11-n standard for the wireless interfaces, and the variant of the power talk introduced in Section 4 with $T_S = 2.5$ ms, $Q = 8$ and $\gamma = 0.25$ V. In conclusion, placing power talk side by side with the high bandwidth, but at the same time unreliable, wireless network, and exploiting the cyber-physical properties of the MG, is a viable and promising solution to increase the system robustness.

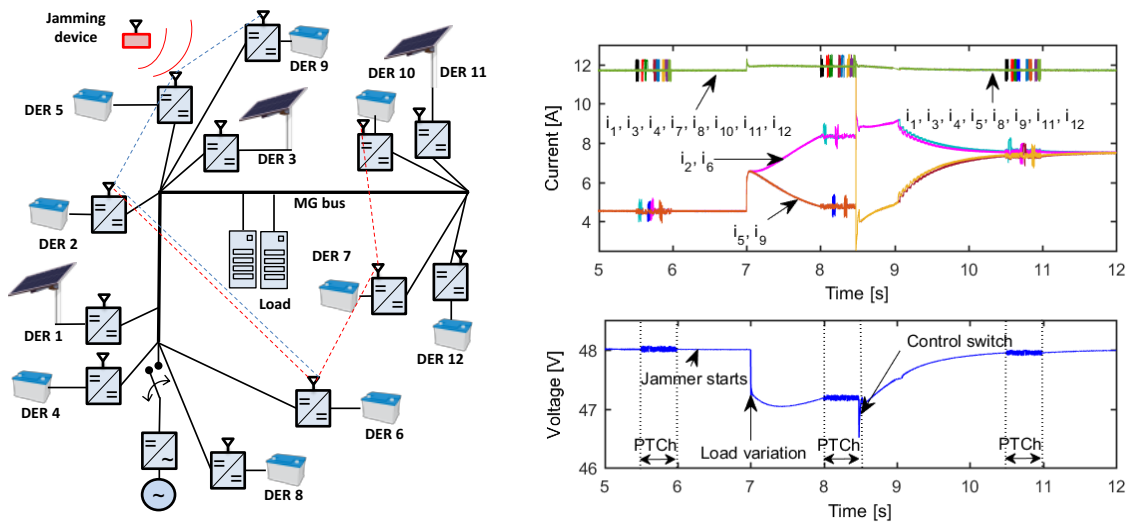

*Figure 6: (a) The MG considered in the study case. The dashed blue lines represent the links before the reconfiguration, the dashed red the communication graph after the activation of the jammer (located in the upper-left corner). (b) PLECS simulation of the secondary control reconfiguration in case of jamming attack. We report the output current of each DER and the voltage measured on the MG bus. Observe the periodic power talk channels (PTCh) used to signal the network information.*

## 6. CONCLUSIONS

This paper has reviewed the use of power talk in DC microgrids, a low-rate communication technique that reuses the power electronics and does not rely on a dedicated communications hardware. The use of power talk has both architectural and functional implications on the operation of system of MGs and can support multiple smart energy applications. Despite the low rate, the reliability and the security of the low-rate communication channel offered by power talk can have a significant impact on the overall performance. We have presented two case studies that clearly show the utility and the potential of power talk, one related to economic dispatch and the second to the cybersecurity in MGs. Future work includes integration of power talk in other smart energy applications and processes and co-design of power talk with the distributed control algorithms.